\documentclass[11pt, executivepaper]{article}
\usepackage[utf8]{inputenc}
\usepackage[T1]{fontenc}
\usepackage{natbib}
\usepackage{amsmath, amsthm, amssymb}
\usepackage{amsfonts}
\usepackage{mathtools}
\usepackage{xcolor}
\usepackage{graphicx}
\usepackage{enumitem}
\usepackage{geometry}
\usepackage{hyperref}
\hypersetup{colorlinks= true, allcolors=blue}
\setcitestyle{aysep={}}

\begin{document}

\title{\emph{Orthodox or Dissident?
\\
The Evolution of Bohm’s Ontological Reflections in the 1950s}}

\author{Andrea Oldofredi\thanks{Contact Information: Centre of Philosophy, University of Lisbon, Alameda da Universidade 1600-214, Lisbon, Portugal. E-mail: aoldofredi@letras.ulisboa.pt}}

\maketitle

\begin{abstract}
David Bohm has often been considered unable to understand the meaning of the quantum revolution as well as its radical metaphysical implications. Similarly, his pilot-wave theory was negatively portrayed as an attempt to restore a classical and deterministic \emph{Weltanschauung}. Against this background, the aim of this paper is twofold: in the first place, it will be argued that the accusations of dogmatism advanced by several eminent physicists \emph{contra} Bohm show a biased understanding of his works. Referring to this, two case studies will be discussed: the Bohm-Pauli correspondence, and the difficult relationship between the former and Leon Rosenfeld, a fervent supporter of Bohr's philosophy of complementarity. These examples indicate that the opposition to the pilot-wave approach was for the most part not based on scientific grounds. In the second place, I will reconstruct and analyze the evolution of Bohm's philosophical reflections about ontology, scientific realism and pluralism studying private correspondences as well as his main works in the fifties culminated in the book \emph{Causality and Chance in Modern Physics}. Underlining the originality of Bohm's thoughts, it will be concluded that his perspective can be characterized as a form of internal realism. 

\vspace{4mm}

\end{abstract}
\vspace{5mm}

\begin{center}
\emph{Accepted for publication in The European Physical Journal H}
\end{center}

\vspace{10mm}

\noindent \textbf{Acknowledgements:} I acknowledge the support of the Birkbeck College, University of London for their kind permission to publish letters of David Bohm whose originals are held in their archives, and for having provided an electronic copy of the archival material. Many thanks go to the audiences in Warsaw and Lisbon for valuable feedback on previous drafts of this essay as well as to Olga Sarno, Marij van Strien and the two anonymous referees of \emph{The European Physical Journal H} for careful comments on the final version of the manuscript. Financial support for this research has been provided by the Funda\c{c}$\tilde{\mathrm{a}}$o para a Ci\^encia e a Tecnologia (FCT) (Grant no. 2020.02858.CEECIND).
\clearpage

\tableofcontents
\vspace{5mm}

\section{Introduction}

Quantum Mechanics (QM) is one of our most accurate answers to questions concerning the intrinsic structure of matter. Such a theory describes the behavior of elementary particles, atoms and molecules in a way that drastically changed our classical conception of the world, speaking about ontological indeterminacy and inherently stochastic quantum jumps. For instance, according to its standard formulation isolated quantum systems do not instantiate properties with definite values, i.e.\ quantum items possess indefinite attributes prior to experimental observation, marking a significant ontological difference with respect to classical objects. Measurements in QM, then, do not reveal pre-existing values of physical magnitudes---which in fact are essentially contextual---and their results are inherently probabilistic. In this theoretical context it is therefore a meaningless effort to look for a causal story that would explain a particular experimental outcome from its initial conditions (cf.\ \cite{Dirac:1947}, \cite{Heisenberg:1949}, \cite{vonNeumann:1955aa} \cite{Sakurai1994}). This particular image of the world found large support and endorsement among the physicists who contributed to the quantum revolution---as for example Bohr, Born, Dirac, Heisenberg, Jordan and Pauli to mention a few---to the point that such a \emph{Weltanschauung} soon became the orthodoxy with respect to the interpretation of the quantum formalism.

Although historians and philosophers of physics argued that it is disputable whether a common metaphysical perspective existed among the fathers of quantum mechanics\footnote{There was disagreement for example about whether the wave function undergoes an actual collapse in measurement situations. Whereas according to Bohr there was no such a collapse of the $\psi$ function, being entanglement and complementarity the real novelties introduced by QM  (cf.\ \cite{Howard:2004aa}, p.\ 672), for Born, Dirac, Heisenberg, von Neumann and others the stochastic jumps and the non-commutativity of quantum operators were the primary innovations of quantum theory.}, there is however a precise sense in which one may properly speak about a cohesive and unitary orthodox or ``Copenhagen'' view of the theory. In fact, Bohr and the physicists who shaped modern QM shared a set of ideas which can be characterized as the inner core of such orthodoxy. Referring to this, Freire underlined that 
\begin{quote}
in spite of the existence of important differences, both the intellectual backgrounds and the scientific views of people like Bohr, Pauli, Heisenberg, Born, and Jordan, who had been working together on the collective construction of quantum mechanics, had several points in common. All of them endorsed both indeterminism and the assumption of the corpuscular and discrete nature of atomic phenomena. They also firmly believed in the completeness of quantum theory. [...]. [They] were attached to the revolutionary character of quantum mechanics, and were unsympathetic to any attempt to restore such classical ideals like causality and visualizability in microphysics (\cite{Freire:2015aa}, p.\ 81). 
\end{quote}

\noindent On the same vein, \cite{Osnaghi:2009} note that:
\begin{quote}
[t]he existence of an ``orthodox view'' of quantum mechanics was generally taken for granted since the 1930s. However, the meaning of such a label was far from being univocally determined. Several factors contributed to keeping its definition vague, and by the same token to reinforcing the impression that an orthodox view did indeed exist (\cite{Osnaghi:2009}, p.\ 99).
\end{quote}

\noindent Remarkably, such an orthodox view was so vastly supported by the founding fathers of QM that they not only formed an intellectual imperialism---as argued by Heilbron (\cite{Heilbron:1988}, p.\ 202)---but also believed that their philosophical perspective constituted the only feasible interpretation of the quantum formalism, and not merely a possible reading of it (\cite{Beller:1999}, p.\ 191).

Against this background, since the birth of quantum theory until the 1950s there were a few eminent physicists dissatisfied by the philosophical content of quantum theory. For instance, on the one hand Einstein, Podolsky and Rosen famously argued that the QM could not have been considered a complete description of the physics at quantum scales, opening the discussion about its possible completion with hidden parameters (cf.\ \cite{EPR:1935}, see also \cite{Einstein:1949, Einstein:1953}). Notably, Einstein endorsed a statistical (or ensemble) interpretation of the quantum formalism, showing in several places his discomfort with the usual view. On the other hand, Schr\"odinger proposed a realist interpretation of $|\psi|^2$ as charge density, and later proved that the measurement problem is a logical consequence of the principles of QM (cf.\ \cite{Schrodinger:1935}). Despite these criticisms against the orthodoxy, Schr\"odinger's proposal was shown to be empirically inadequate, while Einstein's opposition to such philosophy of QM never translated into a full-fledged interpretation of the theory. 

The very first alternative reading of QM, in fact, saw the light only in the early fifties with the work of David Bohm, who rediscovered and extended de Broglie's pilot-wave theory (cf.\ \cite{deBroglie:1927}; for a historical account see \cite{Freire:2015aa, Freire:2019}). It is well-known, however, that such an interpretative framework was poorly received in virtue of its ontological picture, which was erroneously intended as an attempt to restore a classical---therefore outdated---worldview (cf.\ \cite{Freire:2005, Freire:2015aa}, \cite{Passon:2018}, \cite{vanStrien:2019a}), gaining new momentum only after John Stuart Bell's interest in it, praising the value of Bohm's ideas, as recently argued in \cite{Freire:2024}. 
On the contrary, as we will see in the remainder of this work, Bohm's goal was to show that QM could have been provided with a clear ontology. Therefore, the indeterminate and fuzzy aspects of the orthodox view were not necessary features of the theory and could have been removed from its metaphysics, reintroducing an \emph{anschaulich}, i.e.\ visualizable, intelligible picture of the physical processes taking place at quantum length scales.\footnote{The debate between Heisenberg and Schr\"odinger on the notion of \emph{Anschaulichkeit} in quantum theory is nicely resumed in \cite{Uffink:2016}.}

The aim of this paper is twofold: in the first place, I will explain that the accusations of dogmatism made by many physicists \emph{contra} Bohm are for the most part scientifically unfounded, showing a philosophically biased understanding of the implications and significance of his works (Section \ref{Pauli}). Referring to this, after a brief introduction of the pilot-wave theory, in the next section two important case studies are discussed: the Bohm-Pauli correspondence and the difficult relationship between Bohm and Leon Rosenfeld, who was a fervent supporter of Bohr's philosophy of quantum mechanics. As the reader will see, both examples clearly indicate that the opposition against the pilot-wave theory was generally not based on scientific grounds, but rather were of conceptual nature. In the second place, I will reconstruct and analyze the evolution of Bohm's philosophical reflections about ontology, scientific realism and pluralism studying private correspondences as well as his main works in the fifties culminated in the book \emph{Causality and Chance in Modern Physics}, unfortunately largely forgotten in contemporary discussions (Section \ref{50s}). Underlining the originality of Bohm's thoughts, it will be argued that his perspective can be characterized as a form of internal realism (Section \ref{IR}).\footnote{NB: no connection with Putnam's version of internal realism is meant in the present essay.} Conclusions are drawn in Section \ref{conc}.

\section{David Bohm: Orthodox or Dissident?}
\label{Pauli}

David Bohm joined the foundational debate proposing the very first alternative interpretation of the quantum formalism in his papers \emph{A suggested interpretation of the quantum theory in terms of hidden variables Part I \& II} published in January 1952 in the prestigious \emph{Physical Review}.\footnote{In 1951 Bohm published the textbook \emph{Quantum Theory}, providing an introduction to the standard interpretation of QM. His interests in the foundational debates and his dissatisfaction with the traditional viewpoint was sparked (among other sources) by several discussions with Einstein, who is also acknowledged in Bohm's 1952 papers. For details on the political and personal vicissitude of Bohm between 1950 and 1952 see \cite{Freire:2005, Freire:2015aa, Freire:2019}.} There he showed that a causal description of quantum phenomena in terms of particles moving along continuous trajectories was not only possible, but also mathematically and physically consistent. 

In his essays Bohm criticized what he called the ``usual interpretation'' of QM---referring mainly to Bohr's and Heisenberg's ideas---objecting that the mere empirical consistency of quantum theory is not a sufficient reason to exclude other (possibly better) alternative interpretations. Similarly, he disputed the popular view according to which QM forces us to abandon the idea of a precise characterization of physical systems at quantum scale as well as an accurate description of their dynamical evolution, since it is not proved that ``such a renunciation is necessary'' (\cite{Bohm:1952aa}, p.\ 168). Moreover, he was deeply dissatisfied by the inability of the standard theory to explain the actualization of macroscopic measurement outcomes. 

To overcome these shortcomings Bohm's interpretation exhibits a clear metaphysics describing quantum systems in terms of particles guided by wave functions---considered real physical fields---whose dynamical evolution is in turn governed by the Schr\"odinger equation
\begin{align}
\label{SE}
i\hbar\frac{\partial }{\partial t}\psi=H\psi.
\end{align}

\noindent Such a proposal is ontologically unambiguous because Bohmian corpuscles always have a precise localization in three-dimensional space and a well-defined velocity independently of any observation. Contrary to the principles of standard QM, then, the postulate according to which $\psi$ alone provides the complete specification of a certain system is rejected. To this regard, Bohm also stated that in his framework the uncertainty principle becomes just ``an effective practical limitation on the possible precision of measurements'' (\cite{Bohm:1952aa}, p.\ 171). Hence, it should not be interpreted as an irreducible impossibility \emph{to conceive} position and momentum as simultaneously defined quantities, in open opposition to what has been claimed in \cite{Heisenberg:1927}. 

In order to outline Bohm's theory one can express the wave function in polar form $\psi=Re^{(iS/\hbar)}$---where $R$ and $S$ represent two coupled real functions corresponding to the amplitude and the phase of the wave. Then, posing $P(x)=R^2(x)$, where $P(x)$ represents the probability density, one obtains the following equations for $R$ and $S$:
\begin{align}
\label{cont1}
\frac{\partial P}{\partial t} + \nabla\cdot\left(P\frac{\nabla S}{m}\right)=0,
\end{align}
\begin{align}
\label{QHM}
\frac{\partial S}{\partial t} + \frac{(\nabla S)^2}{2m}+V(x)-\frac{\hbar^2}{4m}\Bigg[\frac{\nabla^2 P}{P}-\frac{1}{2}\frac{(\nabla P)^2}{P^2}\Bigg]=0.
\end{align}

\noindent The former is the quantum continuity equation for the probability density, whereas the latter is the quantum Hamilton-Jacobi equation describing the motion of a particle (or a configuration of particles) with kinetic energy $(\nabla S)^2/2m$ and subjected to the influence of both a classical and a new quantum potential:
\begin{align}
U(x)=\frac{-\hbar^2}{4m}\Bigg[\frac{\nabla^2 P}{P}-\frac{1}{2}\frac{(\nabla P)^2}{P^2}\Bigg]=\frac{-\hbar^2}{2m}\frac{\nabla^2R}{R}.
\end{align}

\noindent Notably, Bohm claimed that considering an ensemble of particles' trajectories solutions to the above equations \eqref{cont1}-\eqref{QHM}, then ``if all these trajectories are normal to ant given surface of constant $S$, then they are normal to all surfaces of constant $S$, \emph{and $\nabla S/m$ will be equal to the velocity vector $v(x)$ for any particle passing the point $x$}'' (\cite{Bohm:1952aa}, p.\ 170, my emphasis). Therefore, \eqref{cont1} can be rewritten as 
\begin{align}
\label{cont}
\frac{\partial P}{\partial t} + \nabla\cdot (Pv)=0,
\end{align}
\noindent where $Pv$ is interpreted as the mean current of the particles in a given configuration. This equation is particularly important since the Born distribution holds as a consequence of it. Finally, the empirical adequacy of the pilot-wave theory is formally established since it postulates the quantum equilibrium hypothesis, according to which if a certain configuration of particles is $|\psi|^2$-distributed at an arbitrary initial time $t_0$, then such distribution will be maintained at every later time $t$.\footnote{There have been many efforts to \emph{derive} the Born's rule in the pilot-wave theory as for instance \cite{Bohm:1953b}, \cite{Durr:2013aa} Chapter 2, \cite{Valentini:1991aa}. However, the debate about status of the quantum equilibrium hypothesis in Bohm's theory is still open, for an overview cf.\ \cite{Callender:2007aa}.} Referring to this, it should be stressed that in Bohm's theory probabilities do not refer to a inherent indeterminacy of quantum objects, but rather possess an epistemic character: in experimental situations we can neither know, nor manipulate the initial positions of the particles, consequently measurement outcomes will be practically unpredictable. Thus, in this context randomness arises as in classical statistical mechanics. In addition, it is worth mentioning that in the second part of his paper Bohm provides a precise theory of quantum measurement, explaining how experimental results emerge from the motion of individual quantum particles without the introduction of stochastic quantum jumps, filling the explanatory gap left by quantum theory (cf.\ \cite{Bohm:1952ab}, Section 2). This remarkable achievement showed the actual possibility to provide a detailed description of the physical processes causally responsible of measurement outcomes, avoiding the conceptual  and technical difficulties present in standard QM.

Notwithstanding the formal consistency of this proposal and its empirical equivalence with respect to the predictions of QM, the pilot-wave theory was poorly received being seen as a sterile attempt to restore an anachronistic worldview. As a consequence, Bohm was considered a dogmatic physicist stuck with reactionary and ``orthodox''---this time in the sense of classical---ideas, unable to fully embrace the conceptual revolution generated by quantum theory. In the reminder of this section I discuss two relevant case studies where such a negative opinion clearly emerges.

\subsection{The Bohm-Pauli Correspondence}

Taking into account the Bohm-Pauli correspondence occurred between July and December 1951 one can readily understand that the Austrian physicist had little regard of Bohm's work, considering it a plagiarism of de Broglie's pilot-wave theory.\footnote{Notably, Bohm was unaware of de Broglie's work on the pilot-wave theory, Pauli mentioned it to him, cf.\ \cite{Pauli:1996}, p.\ 346.} As it is well-known, Pauli raised strong objections against the latter at the Solvay congress in 1927 (cf.\ \cite{Bacciagaluppi:2009aa}, pp.\ 235-238; the transcription of Pauli's contribution to the discussion about de Broglie's report can be found  at pp.\ 400-401), and from the letters at our disposal one may fairly say that he approached Bohm's proposal with a negative bias. By the summer 1951, in fact, Pauli did not yet read carefully his manuscripts,  dismissing the new formulation of the pilot-wave approach as a ``simple minded'', cheap solution to the problems of QM (cf.\ Letters 1263 and 1264 in \cite{Pauli:1996}). Evidence for these claims can be found also in other letters that Pauli sent to Fierz, Panofsky and Rosenfeld. For instance he wrote ``\emph{Die Sache von Bohm ist beinahe ein Plagiat!}''\footnote{\emph{``The Bohm's thing is almost plagiarism!''}, author's translation from the original German.} to the former on 10 January 1952, similarly a month later  Panfosky was told that Bohm's theory was a plain copy of de Broglie's old works of 1926-1927. Even more explicitly, Pauli defined the new pilot-wave theory as a ``revival of de Broglie's old errors of 1927'' in a letter to Rosenfeld dated 16 March 1952 (cf.\ respectively, \cite{Pauli:1996}, Letter 1340, Letter 1364, and Letter 1386). 

This lack of interest may be explained by three related factors: firstly, in the early 1950s physicists were working on the cogent issues affecting quantum field theory---thus, a more advanced theory with respect to non-relativistic QM. Secondly, already at the Solvay conference Pauli, in agreement with Bohr and against the opinions of de Broglie, Lorentz and Schr\"odinger, argued that a spacetime representation of quantum phenomena is not obtainable in virtue of the polydimensional character of the $\psi$-function (cf.\ \cite{Bacciagaluppi:2009aa}, pp.\ 214-216). Finally, looking at the significant empirical and theoretical progresses of the orthodox view, it seemed that this interpretation was the sole possible understanding of quantum physics, as already stressed in the previous section. Given (i) that Pauli was deeply involved in the research about the quantum theory of fields and quantum electrodynamics since the 1930s (cf.\ \cite{Enz:1973}), and (ii) that he was very close to the Copenhagen view, it can safely be claimed that he considered the pilot-wave approach as a dead program, thereby reading Bohm's papers would have been a waste of time. Indeed, Pauli initially attempted to reject Bohm's proposal with the same criticisms he posed against de Broglie in 1927. This can be straightforwardly inferred since in July 1951 Bohm wrote that in the second draft of his papers \emph{all} the objections against de Broglie's theory are answered in detail, calling his attention and directing him to the relevant sections of the essays as for instance Section 7 of \cite{Bohm:1952aa} and in the second Appendix of \cite{Bohm:1952ab}. Referring to this, Bohm insisted in asking Pauli to read thoroughly the papers before discarding the causal interpretation: 
\begin{quote}
With regard to your questions raised in the letter, they are answered in my ``long'' paper. You really have put one in an impossible position. If I write a paper so ``short'' that you will read it, then I cannot answer all of your objections. If I answer all of your objections, then the paper will be too `long''  for you to read. I really think that it is your duty to read these papers carefully, especially if you wish to carry out your promise of sending me your ``permanent and persistent scientific opposition'' (in \cite{Pauli:1996}, p.\ 346, Letter 1264).
\end{quote}

Similarly, in mid-October 1951 Bohm wrote another long letter replying to the same objections reiterated by Pauli in previous correspondence and explaining several details of the pilot-wave theory, with particular attention to scattering of particles, the theory of measurement, the empirical equivalence between his theory and QM. There Bohm pointed again out that ``[i]n the second version of the paper, these objections are all answered in detail'' and that ``[i]t is difficult for me to answer your objections in detail without simply repeating what is in section 7 of paper I and in the first five or six sections of paper II''  (in \cite{Pauli:1996}, p.\ 389, p.\ 390 respectively, Letter 1290).

From the available letters it is possible to deduce not only that Pauli continued to ignore (or to read superficially) Bohm's manuscripts until the fall 1951, but also that he tried to refute the pilot-wave theory appealing to von Neumann's no-go theorem allegedly proving the impossibility to complete QM with hidden parameters (cf.\ \cite{vonNeumann:1955aa}, Chapter 4). To this regard, Bohm explained that such a result is not in contradiction with his theory, underlining essentially that von Neumann implicitly assumed that ``the hidden variables are only in the observed system and not in the measuring apparatus. On the other hand, in my interpretation, the hidden variables are in \emph{both} the measuring apparatus and the observed system. Moreover, since different apparatus is needed to measure momentum and position, the actual results in each respective type of measurement are determined by different distributions of hidden parameters'' (\cite{Pauli:1996}, p.\ 392, Letter 1290, emphasis in the original). Remarkably, he also emphasized to Pauli that von Neumann himself admitted the logical consistency of the pilot-wave approach.\footnote{This fact is reported by Bohm also in a letter to Melba Phillips in early 1952 (printed in \cite{Talbot:2017}, p.\ 147). For details cf.\ \cite{Bohm:1952ab}, Section 9, p.\ 187, \cite{Bricmont:2016aa},  \cite{Freire:2015aa}, p.\ 47 and \cite{Oldofredi:2018}).}

A few weeks later Bohm wrote another letter to Pauli---dated 20 November 1951---who in the meanwhile read with more attention his papers and gave non-negative feedback on them, as we can understand from the first lines of this correspondence. Such a letter is particularly interesting for our discussion, since there Bohm made non-trivial statements about the possibility to \emph{modify} his own interpretation, showing the willingness to develop and generalize it beyond the energy/length scale where non-relativistic QM is applied. In fact, he claimed that the pilot-wave theory may entail new physics at very short length scales, where it admits extensions which would make the evolution equation for the $\psi$-function non-linear, contrary to the case of standard QM (see also Section 9 of \cite{Bohm:1952aa}). To this regard, Bohm proposed an interesting inductive inference about the form of future theories, underlying that 
\begin{quote}
wherever we have found linear differential equations, we have always found that they are only approximations to non-linear equations (e.g.\ sound, light, equation for heat flow, etc.).\ There is no way to prove that the same is not true for $\psi$ waves; in fact, it seems implausible to me to suppose that even though in all other fields, nature must be described by non-linear equations, there will be one field, viz, quantum theory, where no such considerations will ever be needed. But if the equations are actually non-linear in a higher approximation, then the usual interpretation cannot be the ultimate one. I therefore believe that the practical necessity of restricting the description to a part of the world will not necessarily lead to a limitation on the accuracy of description of that part, but that it does so only under present conditions of experiment, in which we do not know how to take advantage of the richer laws prevailing at a more fundamental level, which would permit us to reduce these disturbances far below limits set by our present incomplete laws'' (\cite{Pauli:1996}, p.\ 430, Letter 1309). 
\end{quote}

\noindent Along the same reasoning and considering very short scales, Bohm offered speculative suggestions speaking about the conceivability of measurement of unlimited precision as well as the existence of superluminal signals implying a modification of the special theory of relativity, since---he says---''we have proof only that relativity in its present form holds for distances much greater than $10^{-13}$ cm'' (\emph{ibid.}, p.\ 431). In addition, he even stated that the causal interpretation would be able to provide a new definition of simultaneity which would differ with respect to the one introduced in general relativity. 

Taking into account the elements contained in this correspondence, Bohm's unconventional ideas concerning the restricted validity of physical theories such as relativity and quantum mechanics appear forcefully. Furthermore, it is already clear that he considered the causal interpretation just one among the possible steps towards a better comprehension of quantum physics and not as a final word about its ontology. Hence, one could have easily seen not only the originality of Bohm's views and his scientific pluralist spirit---that would have emerged more systematically in later works---but also that he was not a scientist with a conservative attitude, being ready to envision new paths and directions for his research in both physics and philosophy. 

Interestingly, Pauli reviewed Bohm's paper for \emph{Physical Review} as underlined in \cite{Talbot:2017}, and on 3 December 1951 he eventually acknowledged the internal consistency of Bohm's work writing that: 
\begin{quote}
I do not see any longer the possibility of any logical contradiction as long as your results agree completely with those of the usual wave mechanics and as long as no means is given to measure the values of your hidden parameters both in the measuring apparatus and in the observed system (\cite{Pauli:1996}, p.\ 436, Letter 1313).
\end{quote}

\noindent However, he also raised two distinct problems concerning the treatment of photons and the lack of description of the phenomenon of pair creation:
\begin{quote}
I wish also to point out that the whole streamline picture is essentially non relativistic (as it fails both for photons and for pair generation). Therefore I cannot consider an argument as sound, which claims to reform the theory in the relativistic region, whilst it attacks just the non relativistic part of the theory which is correct (\emph{ibid.}). 
\end{quote}

\noindent Bohm faced these criticisms in a letter sent at the end of December 1951 closing their correspondence for that year.\footnote{From an exchange between Pauli and Pais we know that between mid-April and the beginning of May 1952 Bohm wrote to the former a ``very crazy and impudent letter'' and they were then in rather bad terms, cf.\ \cite{Pauli:1996}, p.\ 627, Letter 1412. Unfortunately, this letter has been lost.} In order to reply to these last objections, he underlined that in his interpretation photons are not treated as particles. Indeed, as shown in the appendix A of \cite{Bohm:1952ab} field configurations are introduced, corresponding to the transverse part of the vector potentials $A(x)$. In his work, as underlined also in \cite{Struyve:2010aa}, Bohm provides a guidance equation for the field coordinates, and the $\psi$ function is interpreted as a functional of all the the Fourier components of the vector potentials. As Bohm pointed out, his treatment of the electromagnetic field leads to the same prediction of the standard formulation of QM, answering thereby the first criticism.

In order to reply to the second objection the Dirac sea picture is explicitly mentioned for the first time, ``[t]here, the creation of a pair is simply a transition from a negative to a positive energy state''. Such a pilot-wave formulation of the hole theory was developed in the subsequent years, as we can see in \cite{Bohm:1953aa}, reaching a mature formulation in his last published work \cite{Bohm:1993aa}.\footnote{This revival of Dirac's ideas have been recently put forward in many essays as for instance \cite{Colin:2003aa}, \cite{Colin:2007aa} and \cite{Deckert:2019}.} Thus, Bohm concluded, ``there are no precesses which can be treated by the usual interpretation and which cannot also be treated in the causal interpretation'' (\cite{Pauli:1996}, p.\ 444, Letter 1315).

To present historical facts accurately and to be fair with Pauli's criticisms, it should be noted that the Dirac sea hypothesis was poorly considered at that time (cf.\ \cite{Kragh:1990} for details). Thus, Bohm's answer to the question about pair creation could have left him unsatisfied. In addition, it should be underlined that Pauli's last objection is still employed as one of the main arguments against the Bohmian interpretation of the quantum formalism. Nevertheless, we should also say on the one hand that the standard formulation of quantum field theory itself is not immune from severe philosophical issues, and on the other hand that interesting steps have been made in order to provide a coherent pilot-wave approach to the phenomena of particle creation and annihilation, as for instance in \cite{Colin:2003aa}, \cite{Colin:2007aa}, \cite{Durr:2004aa}, \cite{Tumulka:2022}. Thus, although pair creation were not treated by the theory contained in the 1952 papers, this was not a motivation to discard in principle this new research program. Therefore, even in the absence of arguments able to prove the pilot-wave theory definitively flawed and to consider Bohm a reactionary physicist, Pauli nonetheless continued to describe him as a dogmatic thinker and to criticize his approach to QM.\footnote{See also the irreverent, almost disrespectful description of the pilot-wave theory given by Pauli in a letter to Fierz recalling anecdotes of 1927 (\cite{Pauli:1996}, Letter 1337).} For example, he defined Bohm as a ``Sektenpafaff'' (a priest of a sect) in his letters to Fierz, Pais and Stern---cf.\ \cite{Pauli:1996}, Letter 1337, Letter 1412 and Letter 1454 respectively---and his work as ``\emph{ein klassisch-deterministischer Mythos des atomphysikalischen Geschehens}'' (a classical deterministic myth of atomic physics) in his letters to Fierz (\cite{Pauli:1996}, Letter 1337 and Letter 1368).

In the same vein, after the publications of Bohm's papers Pauli felt the need to discredit them publicly for two related reasons. In the first place, he feared that without a public reaction against the causal interpretation his opposition would have been reduced to a mere philosophical disagreement---his worries were justified since Bohm communicated e.g.\ to Einstein that Pauli eventually acknowledged the logical consistency of the pilot-wave approach.\footnote{In December 1951 Bohm wrote to Einstein: ``[i]t may interest you to know that Pauli has admitted the logical consistency of my interpretation of the quantum theory in a letter, but he still rejects the philosophy. He states that he does not believe in a theory that permits us even to $\underline{\textrm{conceive}}$ of a distinction between the observer's brain and the rest of the world''. Folder C11, David Bohm Papers, Birkbeck College, University of London.} To this regard he wrote to Fierz:
\begin{quote}
There is also the danger that---if I simply remain silent---Bohm will spread the word that I have nothing to object to his ``theory'' ``except philosophical prejudices'' (\cite{Pauli:1996}, p.\ 501, Letter 1337, author's translation from the original German).
\end{quote}

\noindent Similar concerns appear also in a correspondence to Rosenfeld:
\begin{quote}
It was necessary for me to write something about it, because I am not only always asked ``what I think about it'', but also because the younger fellow travellers of Bohm (mostly `deterministic' fanaticists, more or less marxistically coloured) are spreading incorrect rumors about my opinions. (They also try to persuade de Broglie, that there is some truth in his old attempts of 1927) (\cite{Pauli:1996}, p.\ 582, Letter 1386).
\end{quote}

In the second place, Pauli was afraid that Bohm's theory would have found support among young scientists, especially in France where Bohr's complementarity doctrine was negatively received\footnote{This can be understood from Pauli correspondence with Destouches in \cite{Pauli:1996}.}, as we can understand by the already mentioned letter to Fierz, (cf.\ Letter 1337 in \cite{Pauli:1996}). In fact, in his essay honoring the 60th birthday of de Broglie, Pauli wrote an essay that criticized explicitly the pilot-wave theory, arguing that it did not preserved the symmetry between position and momentum representation. Moreover, he did not waste occasion to shared his negative opinions about such a theory in several letters warning other physicist not to pay attention to it as reported in \cite{Freire:2005}. 

In sum, Pauli was and remained convinced that Bohm had a dogmatic faith in determinism, portraying him as a physicist with an obsolete \emph{Weltanschauung} anchored to outdated ideas.\footnote{It is interesting to highlight an ironic plot-twist: in one of their last exchanges, the ``reactionary'' Bohm accused Pauli to be excessively conservative, being stuck with old positivistic ideals that prevented him to appreciate the novelty of his proposal:
\begin{quote}
Since you admit the logical consistency of my point of view, and since you cannot give any arguments showing that it is wrong, it seems to me that your desire to hold on to the usual interpretation can have only one justification; namely, the positivist principle of not postulating constructs that do not correspond to things that can not be observed. This is exactly the principle which caused Mach to reject the reality of atoms, for example, since no one in his day knew how to observe them. [...] After all, we must not expect the world at the atomic level to be a precise copy of our large scale experience (as proponents of the usual interpretation are so fond of saying). Rather than accept a perfectly logical and definite concept of polydimensional reality that leads to the right results in all known cases, and opens up new mathematical possibilities, you prefer the much more outlandish idea that there is no way to conceive of reality at all at the atomic level. Instead you are willing to restrict your conceptions to results that can be observed at the long scale level, even though more detached conceptions are available, which show at least, never the production of these results might be understood causally and continuously (\cite{Pauli:1996}, p.\ 442, Letter 1314).
\end{quote}
} 
Given the textual evidence emerging from their correspondence and from Bohm's published papers, however, this opinion seems to be rather unjustified. On the one hand, determinism was not an issue for Bohm (cf.\ \cite{vanStrien:2024} and \cite{DelSanto:2023}); he thought in fact that new quantum theories will be needed in order to describe the physics at very short length scales, and he expected them to be non-linear for the $\psi$-field (cf.\ Letter 1309 in \cite{Pauli:1996}). On the other hand, in his letters Bohm provided innovative views about several other issues, as for instance the poly-dimensional character of $\psi$, the non-locality of his approach and the limited validity of our best physical theories, showing his pluralist attitude about interpretational debates. 

\subsection{Rosenfeld's War Against Bohm's ``Obscurantism''}
 
Leaving behind the Pauli's case---that shows a negatively biased reception of the pilot-wave theory---it is worth noting that Bohm's work has often been discarded and discredited by several physicists without solid scientific motivations, frequently on political grounds. A remarkable example is given by Oppenheimer's reaction to the causal interpretation.\footnote{Oppenheimer was Bohm's PhD advisor.} As the physicist Max Dresden reported in an interview (see \cite{Peat:1997}, p.\ 133), during a talk he gave in Princeton---the former affiliation of Bohm before his emigration from U.S.\ to Brazil---the audience's reaction was emblematic: no one read Bohm's paper since it was simply considered a waste of time. Oppenheimer declared that his theory was a form of ``juvenile deviationism'', an expression used also by Abraham Pais. Remarkably, at the end of Dresden's talk Oppenheimer sentenced that ``if we cannot disprove Bohm, then we must agree to ignore him'' as reported also by Bricmont (\cite{Bricmont:2017}, p.\ 203). 

Against this hostile background, the second historical case study that we will be considering concerns the harsh ideological critique of Bohm's theory made by L\'eon Rosenfeld, who was ``Niels Bohr's closest assistant for epistemological matters'', as underlined by Freire. He was himself a Marxist, and being very close to the philosophy of complementarity ``saw the battle against the causal interpretation as part of the defense of what he considered to be the right relationship between Marxism and science'' (\cite{Freire:2015aa}, p,\ 36). 

As we have already seen a few lines above, Rosenfeld and Pauli exchanged letters in which they spoke about Bohm's reformulation of the pilot-wave theory. From the correspondence of 16 March 1952 Rosenfeld's negative opinion about the latter is crystal-clear:
\begin{quote}
I hope that the people who told you that I was ``interested'' in Bohm's heresy did not suggest that I was in any way impressed by it! I am only interested in stamping out this new obscurantism, because it is positively harmful; I know some of Bohm's ``fellow-travellers'' and am distressed to see such intelligent and sincere young people waste their energy in this way (\cite{Pauli:1996}, p.\ 587, Letter 1389).
\end{quote} 

\noindent Apart from the colorful language used against Bohm's proposal, this letter is significant for two other reasons. On the one hand, Rosenfeld explicitly shared Pauli's need to publicly attack the pilot-wave theory in order to dissuade young scientists from supporting this interpretation. To this regard he wrote: ``I feel we have also a duty to help these people out of the bog if we can. Your article is very forceful indeed and I enjoyed it very much; I hope it will make due impression'' (\emph{ibid.}). On the other hand, and contrary to Pauli's strategy, Rosenfeld thought that the causal interpretation must be fought on philosophical rather than physical grounds, since its metaphysics contains ``the root of all evil''. Because of Bohm's explicit criticism to Bohr's view in his 1952 paper, Rosenfeld aimed at showing the inconsistency of the ``metaphysical character of the deterministic pseudo-interpretation of quantum theory''.

Once again the issue of determinism---the source of Bohm's ``obscurantism''---is mentioned, regardless of Bohm's own arguments according to which the pilot-wave theory (i) is derived from the mathematical structures of quantum theory, and (ii) may be modified and made non-linear to describe physics at very short length scales of order 10$^{-13}$m, as clearly stated in Section 9 of \cite{Bohm:1952aa}. 

It is worth noting that in May 1952 Rosenfeld personally wrote to Bohm in order to explain his reasons to deny the very existence of a debate concerning the interpretation of quantum theory. As reported by Freire, in a letter dated 30 May 1952 we read: ``I certainly shall not enter into any controversy with you or anybody else on the subject of complementarity, for the simple reason that there is not the slightest controversial point about it'' (\cite{Freire:2015aa}, p,\ 36). This is another clear illustration of what we already said above, namely that the Copenhagen perspective---in this case the Belgian physicist was referring to Bohr's complementarity doctrine---was so widespread and rooted that it was the only conceived interpretation of quantum theory. 

Referring to this, it is interesting to underline that Rosenfeld supported a different 
Marxist approach with respect to Bohm, he in fact rejected determinism, as many other Marxist physicists did, as for instance Vladimir Fock, Paul Langevin or Boris Hessen to mention a few. Thus, Rosenfeld viewed the deterministic character of the pilot-wave theory as a return to surpassed classical ideas, contrary to the principle of complementarity which entails the abandonment of determinism, and that can be understood in a dialectical manner as noted by \cite{vanStrien:2024}.\footnote{For details on Marxist approaches to quantum theory cf.\ \cite{Freire:1997}.} 

This criticism related to the deterministic character of Bohm's theory was shared also by other physicists as for instance Heisenberg, who stressed the unsuccessfulness of the then proposed alternatives to QM, in particular he believed that these alternatives tried to ``push new ideas into an old system of concepts belonging to an earlier philosophy'' (\cite{Heisenberg:1955}, p.\ 23). As we have seen, this view is also supported by Pauli who in December 1954 wrote to Born: 
\begin{quote}
Contrary to all reactionary efforts (Bohm, Schr\"odinger etc.\ and in a certain sense also Einstein), I am certain that the statistical character of the $\psi$-function and thereby of the laws of nature (...) will determine the style of the laws for at least a few centuries. It could be that later on, something completely new will be found, for example in connection with the processes of life; but to dream of a way back, back to the classical style of Newton-Maxwell (and they are merely dreams, to which these gentlemen dedicate themselves) seems to me without hope, devious, of bad taste. And, we may add, it is not even a good dream (\cite{Pauli:1999}, p.\ 887).
\end{quote}
 
\noindent Notably, not only Rosenfeld attacked Bohm's theory in papers and workshops---particularly important is the Colston Symposium held in Bristol in 1957, where he criticized publicly the causal approach claiming that the complementarity view was the only conceivable option for QM (for more details see \cite{vanStrien:2019}, Section 1.2)---but also invited colleagues to oppose this proposal and to prevent its diffusion and circulation:
\begin{quote}
[h]e pushed Fr\'ed\'eric Joliot-Curie---a Nobel prize winner and member of the French Communist Party---to oppose French Marxist critics of complementarity; advised Pauline Yates---Secretary of the ``Society for cultural relations between the peoples of the British Commonwealth and the USSR''---to withdraw her translation of a paper by Yakov Ilich Frenkel critical of complementarity from Nature (\cite{Freire:2015aa}, p,\ 38).
\end{quote}

\noindent Juliot-Curie did not take part to Rosenfeld's campaign, but several distinguished physicists adhered to it, as for instance Abraham Pais, Vladimir Fock and Adolf Gr\"unbaum among others. Notably Fock, one of the most prominent figure of USSR physics, called the pilot-wave theory `a widespread illness', thinking that such an approach would represent a dead-research program given its return to classical ideas (cf.\ \cite{Fock:1957}). Finally, to mention one last result of Rosenfeld's efforts against the diffusion of the causal view, he ``asked Nature not to publish a paper by Bohm entitled ``A causal and continuous interpretation of the quantum theory'', and advised publishers not to translate one of de Broglie's books dedicated to the causal interpretation into English'' (\cite{Freire:2015aa}, p,\ 38).

As we have seen with many historical examples, several criticisms against the pilot-wave theory were not based on scientific grounds, and numerous physicists were opposed to this approach primarily for ideological or philosophical reasons, being convinced that the orthodox view could not be questioned. Notably, it was a shared opinion that Bohm tried to promote an old and outdated worldview. However, his scientific pluralism and non-classical ideas emerge vigorously from his public and private intellectual productions, showing also a strong anti-dogmatic attitude towards the interpretational debate concerning the meaning of the quantum formalism. Referring to this---and contrary to the beliefs held by many critics---he considered the causal approach only an initial attempt to improve the physical and philosophical content of QM. In fact, Bohm himself suggested possible ways to modify the pilot-wave theory in order to test it against ordinary QM at sub-quantum scales, showing the will to modify, extend and revise his own theoretical framework---indeed, he advanced a stochastic approach to the causal view in a later work with the French physicist J.P.\ Vigier (cf.\ \cite{Bohm:1954aa}). Moreover, he was ready to accept the non-classical feature of the theory as the polydimensional character of the $\psi$-field and non-locality. In addition, it is remarkable that many contested the deterministic character of the causal interpretation, reading it as a sign of Bohm's conservative views, while determinism is merely a mathematical feature of the formalism employed. Thus, we can safely say that determinism was neither an a priori assumption of the causal view, nor some characteristic that Bohm wanted to restore.

Unfortunately, notwithstanding the actual content of his papers and letters, he was seen somewhat ironically as both a reactionary physicist whose aim was to reintroduce classical ideas as well as as a dissident for his challenge to the orthodox view. 

If in this section we argued that it is incorrect to portrait him as a conservative scientist, in what follows we will introduce in detail Bohm's reflections on the structure of reality and the representational power of physical theories. As we will see, Bohm's philosophy of physics not only reveals interesting observations about the epistemic role of science, but also it proves useful even for contemporary debates about scientific realism and pluralism.

\section{The Infinite Structure of Reality}
\label{50s}

Contrary to the positivistic faith that David Bohm ascribed to several physicists supporting the orthodox view, he was a realist\footnote{Historical and philosophical investigations show that it is not completely correct to claim that the founding fathers of quantum theory supported purely positivist views. For instance \cite{Howard:1994, Howard:2012}, argue that Bohr's philosophy of quantum mechanics should not be considered positivist or subjectivist, whereas \cite{Oldofredi:2019} show that Dirac was not a physicist guided by a positivistic methodology. Finally, although the young Heisenberg was close to empiricism and positivism, later detached himself from such perspectives, as we can see in \cite{Heisenberg:1958b}.}; namely, he supported the thesis according to which material entities and physical processes exist in the world independently of our minds, knowledge and observations. It is useful to recall here that in the early 1950s he was influenced by Marx and Engels' materialist and naturalist ontology. Consequently, Bohm was convinced that the world is composed only and uniquely by material things which not only exist independently of our minds and knowledge of the world, but also exhaust reality in itself, i.e.\ there are no other supernatural, mental or idealistic entities playing a role in the construction of our universe. Similarly, phenomena and processes that we observe are explained just as interactions between material objects in their various forms.\footnote{For a discussion of the Marxist influence on Bohm's positions about realism in the 1950s cf.\ \cite{Talbot:2017}, Chapter 6, \cite{Freire:2015aa} and \cite{Freire:2019}.} As a consequence of this philosophical position, science should be understood as our best effort to grasp and comprehend the inherent structure of reality, which in turn should be clearly reflected in the principles of physical theorizing.

Referring to this, Bohm had little consideration of positivism---viewed as the ``principle of not postulating constructs that do not correspond to things that can not be observed'', i.e.\ the methodological rule of not postulating the existence of unobservable entities in scientific theories and denying them any ontological commitment---defining it in several places a poor working hypothesis (e.g., in his correspondence with Pauli, \cite{Bohm:1953aa} and \cite{Bohm:1957}).\footnote{Bohm gave the above description of positivism in a letter to Pauli dated mid-December 1951, cf.\ footnote 14 for full bibliographic details. In this correspondence he mentions Ernst Mach as a physicist misguided by positivism although without directly engaging with his philosophical positions. It should be said in addition that in his critical assessment of positivism Bohm generally did refer neither to classic works on this view, as for instance the writings of Auguste Comte, nor to its modern reformulation due to the logical empiricists as e.g., Carnap, Schlick, Reichenbach, Nagel \emph{etc.}. Hence, his definition of positivism may result intuitive and/or na\"ive to contemporary readers.} The history of physics, he stated, shows that the postulation of yet-unobserved entities have often been fruitful for future discoveries and observations---the conjecture about the atomistic nature of matter made in XIX century being one of the most notable examples. Thus, to claim that only what can be (in principle) observed deserves to be considered real is an unjustified limitation to the scientist's ability to provide a precise representation of the physical world. Sticking to such a belief, Bohm argued, is what led generations of physicists to accept the obscure metaphysics of quantum theory as the only meaningful way to interpret its formalism. In a nutshell, this is the reason for which Bohm was not only deeply unsatisfied with Bohr and Heisenberg ideas about QM, as we have seen at the outset of the previous section, but also with the claim that the orthodox interpretation was the only feasible one, as explicitly claimed by Rosenfeld. 

In open opposition w.r.t.\ the then-dominant philosophy of QM, Bohm was convinced that ontological clarity had to be an essential feature of any theoretical framework, and insisted that also quantum theory must have been provided with a clear metaphysical content, as one can easily see from the 1952 papers. As we already mentioned, he argued that the empirical robustness of QM and its contingent mathematical structure are not sufficient motivations to exclude a priori other ontologically clearer formulations of the theory, giving an explicit counterexample to the metaphysical imperialism of the orthodox view.

In addition to ontological accuracy, in order to understand Bohm's philosophical views in the 1950s it is worth considering another key element concerned with the representational power of physical theories---their capacity to describe (portions of) reality. It is indeed crucial to emphasize that in his opinion theoretical frameworks always have a limited domain of application: laws and entities of a given theory can provide a good description of the physical world only within a specific interval of energy/length scales. In turn, reality is characterized in its totality by different levels of description, and no theory can be successfully employed at every scale. As already stressed, Bohm indeed argued the pilot-wave theory as well as standard QM have a limited validity and that ``at distances of the order of 10$^{-13}$cm or smaller and for times of the order of this distance divided by the velocity of light or smaller, present theories become so inadequate that it is generally believed that they are probably not applicable'' (\cite{Bohm:1952aa}, footnote 6). At these regimes he expected that new ontologies and new theories will be discovered, so that the pilot-wave theory and QM would emerge as limiting cases at distance much greater than 10$^{-13}$cm. 

Regarding the finite applicability of quantum theory, Bohm interestingly argued that the mathematical structure of standard QM is itself a limitation in order to find deeper physical laws, and consequently a more precise description of reality. Namely, he claimed both in private correspondence and publicly that quantum theory is based on the unjustified assumption of linearity, which is a central, essential feature for its formalism and cause of many paradoxes (\cite{Schrodinger:1935}, \cite{Maudlin:1995aa})\footnote{Interestingly, Bohm and Pauli discussed the necessity of the linearity of the Schr\"odinger's equation, the reader may refer to Letters 1313, 1314 and 1315 included in \cite{Pauli:1996}. The technical details of such a debate are not strictly relevant to our purposes and will not be mentioned in what follows. A public comment about linearity can be found in \cite{Bohm:1952aa}, Section 9, p.\ 179.}: 
\begin{quote}
among all the possible changes that have been considered, people have avoided questioning the weakest assumption of all; viz, that all of physics must be contained in the theory of linear Hilbert spaces (\cite{Pauli:1996}, Letter 1309, p.\ 430).
\end{quote}

\noindent Speaking about the linearity of the Schr\"odinger's equation, as we have already seen in the previous section, Bohm underlined that in the history of physics as soon as differential equation have been found for a certain law, theory of phenomena, then they would have appeared as approximations to non-linear equations.\footnote{Another important aspect that should be highlighted is that Bohm's justifies his arguments to Pauli with inductive reasons; as we will see below this method by induction will appear frequently in Bohm's philosophy.} Thus, he claimed that there is no procedure to prove that this will \emph{not} happen with the quantum mechanical $\psi$ field as well. Contrary to the standard quantum mechanical case, Bohm underlined to Pauli that

\begin{quote}
the interpretation of de Broglie (as extended by me) is potentially capable of leading to a richer variety of laws of nature than those which are consistent with the usual interpretation. Thus, the usual interpretation must assume a linear and homogeneous equation governing the coefficients of the ``state vectors'' in Hilbert space. From my point of view, no such an assumption is necessary. [...] I believe that the equation governing $\psi$ may, for example be non-linear, and that the usual wave equation is only a linear approximation. [...] (\emph{ibid.}, p.\ 429).
\end{quote}

In particular, in this letter and in Section 9 of \cite{Bohm:1952aa}, he hypothesized (i) that at distances of the order of 10$^{-13}$cm or smaller the wave field will be governed by a non-linear equation of motion, and (ii) the existence of a direct coupling between the $\psi$ field and the hidden parameters in a way that the disturbance induced by an observer in measurement scenarios will be less than those imposed by the Heisenberg uncertainty relations. Thus, Bohm wrote to Pauli ``it is quite possible to contemplate theories which would be inconsistent with the usual interpretation. [...] If this is true, then after we understand the new laws governing short distances, we should be able to make measurements much more precise than would be consistent with the uncertainty principle. Even though the observer still disturbs the hidden variables in the measuring apparatus, the effects of this disturbance on the nuclear system of interest can in principle be reduced a great deal below the limits set by the uncertainty principle, provided that, for example, the equations become appreciably non-linear at short distances'' (\emph{ibid.}, p.\ 430).

These deeper non-linear laws applied at very short distances will allow for a precise description of matter beyond the threshold set by current quantum theory, envisioning experiments with unlimited predictions able to test the predictions of the pilot-wave theory, making it falsifiable. This means that on the one side, the current forms of QM and pilot-wave theory do not constitute the final word about the structure of matter, being valid only in a precise range of energies. On the other side, the possibility to extend the causal interpretation makes it possible to test the hidden variable hypothesis, which if confirmed would have provided a decisive argument against the standard formulation of quantum theory (cf.\ also Section 5 of \cite{Bohm:1953aa}). In Bohm's view this constituted a direct objection against the empiricist basis of QM. 

On the other end of the spectrum, he rejected another perspective, later called by Bohm ``mechanistic philosophy'' (cf.\ \cite{Bohm:1957}), for which reality can be fully explained starting from a fixed set of entities, and a restricted set of laws---something close to what philosophers call foundationalism. He warns us not to expect such a knowledge ``because there are almost certainly more elements in existence than we possibly can be aware of at any particular stage of scientific development.\ Any specified element, however, can in principle ultimately be discovered, but never all of them'' (\cite{Bohm:1952ab}, p.\ 189). This is a hint of the metaphysical infinitism endorsed by Bohm in the 1950s to which we now turn.\footnote{It is interesting to note that Bohm's view about the infinite structure of reality is influenced by Engels' \emph{Dialectics of Nature}, as underlined in \cite{Talbot:2017}, Chapter 6.}

\subsection{The seeds of Metaphysical Infinitism}

To the knowledge of the present author, the very first exposition of Bohm's infinitism has to be found in \cite{Bohm:1953aa}, a paper containing replies to the objections \emph{contra} the causal interpretation made by the Japanese theoretical physicist Takehiko Takabayasi. In this essay Bohm explores a variety of issues among which the possibility to extend the pilot-wave picture to spin and quantum fields, the reality of hidden parameters, the features of the $\psi$-field and the relation between his theory and classical physics.

For the purposes of our discussion it is relevant to mention that in this essay Bohm illustrates the aims and scope of the causal approach even more explicitly with respect to his 1952 papers. In section 4 in fact he states that the main goal of the pilot-wave approach is to show that a logically consistent, causal formulation of quantum mechanics is actually obtainable, and thereby the orthodox view should not be considered the only conceivable interpretation of the quantum formalism. Moreover, Bohm repeats with vigor that his proposal is by no means an effort to provide a final or fundamental ontology for quantum theory---as he stressed several times in his correspondence with Pauli---since there are unlimited possibilities to extend and modify it. He therefore emphasizes again the limited validity and applicability of his proposal. 

Related to this issue, in section 6 of the paper under consideration Bohm provides an interesting discussion concerning the abandonment of causality in the realm of quantum physics. Contrary to the orthodox view---according to which the empirical successes of QM are to be found in the ``renunciation of causality''---Bohm claims that the predictions of the theory are derived from the Schr\"odinger \& Dirac equations together with Born's statistical interpretation of $|\psi|^2$, which can be provided with a causal interpretation. Thus, he argues, the a-causal philosophy of QM is not the key to understand the empirical successes of the theory. In turn, this fact entails that causality can be maintained in quantum domains and that this notion can play a significant explanatory role. However, Bohm sees a potential objection---implicitly present also in Takabayasi's criticisms---namely that preserving some form of causality in quantum theories would represent a return to a classical, Newtonian type of mechanics.

Analyzing this specific issue, Bohm acknowledges that an ``\emph{unlimited} extension of causal laws of the type appearing in classical mechanics would lead to most implausible results'' (\cite{Bohm:1953aa}, p.\ 285). Nonetheless, one should not radically conclude that the concept of causality has to be rejected altogether given the empirical inadequacy of Newton's theory in more fundamental domains. Indeed, such implausible results would emerge not because classical mechanics is inadequate or wrong \emph{tout court}, but rather from an unjustified assumption of universal validity of this theory and its laws. These difficulties would disappear as soon as one admits the restricted validity, and thereby applicability, of Newtonian laws in the classical domain.\footnote{Notably, similar points were raised also in the conclusions of the 1952 papers, where Bohm underlined that ``our epistemology is determined to a large extent by the existing theory. It is therefore not wise to specify the possible forms of future theories in terms of purely epistemological limitations deduced from existing theories'' (\cite{Bohm:1952ab}, p.\ 188). Bohm wrote this sentence in relation to the limited applicability of physical theories, explaining that what is observed depends on the theory at hand---following \cite{Einstein:1936aa}---and therefore it is a dangerous move to extend the laws, concepts and their normative power to domains outside the scope of validity of a certain theoretical framework.}

Interestingly, Bohm underlines that in order to establish whether a Newtonian-type of law would be valid at deeper levels, one would have to analyze rigorously such more fundamental layers of reality and verify ``to what extent a simple theory resembling Newton's laws of motion may be valid there''. The consistency and empirical adequacy of the pilot-wave approach then suggests that causal types of laws are possible even in some quantum domain. Therefore, it is not a priori impossible to extend the notion of causality to---and hence to find causal laws in---the quantum realm: ``[f]or there is nothing intrinsically wrong with classical types of laws, as long as we do not try to extrapolate them unjustifiably by imagining (with Laplace) that they furnish a final theory, or at least a final general framework, within which the details only remain to be filled in'' (\emph{ibid}).

To express this latter point even more clearly Bohm resorted once again to the history of physics. He argued that every time a final theory or a final truth was thought to be achieved, new and more complex levels of reality, phenomena and entities were discovered overturning those conclusions based on an improper universalization of some laws, which were instead appropriate only in specific domains. The transition between classical and quantum physics provides a useful example.\footnote{As Bohm correctly highlights ``conclusions drawn only within the limited domain of the previous laws were however never overturned'', \cite{Bohm:1953aa}, p.\ 286.} Taking historical evidence seriously into account, then, one should not believe that with quantum mechanics or quantum field theory we achieved a final theory of matter. On the contrary, past scientific theories indicate that it is unlikely that we will ever achieve experimental evidence for a framework valid at all levels. Hence, says Bohm
\begin{quote}
it is necessary to formulate our theories in such a way that we explicitly recognize the possibility of an inexhaustible number of new levels, in which entirely new types of laws may be needed. If we do this, then even if we discover that simple Newtonian types of laws do hold in the domain of 10$^{-13}$cm, we know that the final course of the world is not necessarily determined ``mechanically'' by such laws. For there is a continual interaction between all levels; and the more complex laws that may be appropriate to the unlimited number of new levels (which we have hardly even begun to scratch) could easily invalidate the conclusions coming from the unfounded extrapolation of Newtonian laws to \emph{all} levels. Thus, the unsatisfactory aspects of Newtonian types of laws are not present in a theory that limits itself to a finite domain, in which it might hope to verify such laws; but are present only when we try to fit all possible future human knowledge into the limited conceptual framework of these laws (\cite{Bohm:1953aa}, p.\ 286).
\end{quote}

\noindent Notably in a footnote to the above quotation Bohm explicitly says that ``below the level of 10$^{-13}$cm probably lies still another level, etc.\ ad infinitum'', which completes the very first published illustration of his views about reality---which discloses an infinity of different layers---and of science itself, whose goal is not to find absolute and universal truths, but rather to find the correct types of laws and entities at every given level. Thus, since Bohm denied that reality has a fundamental bottom ground---thereby rejecting any sort of foundationalism---we can claim that he endorsed a form of metaphysical infinitism.\footnote{More details will be given below. For an interesting discussion of foundationalism and infinitism cf.\ \cite{Tahko:2018}.} 

It is interesting for our discussion to point out that such ideas were already present before the publication of the 1952 papers, as shown  by Bohm's letter to the mathematician Miriam Yevick dated 28 November 1951: ``Another important concept that must be gotten across is that of the infinite number of levels, that must be used in describing the behaviour of matter. Such a point of view automatically prevents us from closing our concepts, at any particular level'' (\cite{Talbot:2017}, p.\ 207). A few months later on 7 January 1952, Bohm sent another letter to her explaining how the diachronic existence of things depends on the motion of infinite layers of reality:
\begin{quote}
How then do we explain the prevalence of change and the transiency of material things? This is done by the notion of endless transformation. The ``things'' at each level, are made up of smaller ``elements'' at a more fundamental level, and it is the motion of these more fundamental elements (not usually directly visible to us, except with the aid of elaborate scientific research) which causes the appearance and disappearance of the ``things'' existing at a higher level. These more fundamental ``elements'' however, cannot be permanent, but must be made up of still more fundamental “elements” \emph{and so on ad infinitum}. Thus, we can see that every ``thing'' that exists may at some time come into existence and later go out of existence, but there is always a deeper level, in terms of which this change can be viewed rationally as a transformation of a more elementary form of matter, which is not itself basically altered in this particular transformation. Nevertheless, no single ``thing'' is uncreatable or indestructible. Only matter as a whole in its infinity of properties and potentialities is eternal (\cite{Talbot:2017}, p.\ 227, emphasis added). 
\end{quote}

\noindent This quote contains a crucial remark, namely it is clearly stated that the existence of an certain object at a given scale depends upon the motion of entities at more fundamental levels. Generalizing this idea, we can say that the entities defined at a precise scale will be ontologically dependent upon the motion of other infinite items living at more fundamental levels. However, despite the ontological dependence of a given level on deeper layers of reality, the reduction to an absolutely primary class of objects and laws is not achievable since in this account there is no such a fundamental substratum. 

The infinitistic view of reality is again illustrated to Yevick in a letter dated 15 February 1952. Here Bohm interestingly refers to his previous work in plasma physics as an influence to his conception of reality. Indeed, he said, the behavior of a given individual object at a certain scale can be described as constrained by a collective motion of substructures present at more deeper levels, so that the particles that we see with present day technology are in fact constituted by aggregates of other items. Inferring inductively the existence of an infinite number of layers, then, each individual object at a particular level will be ``discovered to be collectively conditioned''. From this claim, one draws the general conclusion that (i) the most fundamental individual items of our current science will be discovered to be collectively conditioned by lower level objects and motions, and (ii) that our ``universe cannot be analyzed into a series of components, each of which are the constituents of the next higher level, and each of which determine the higher levels in a purely analytic way. For the higher levels will also always help determine the character of things that may exist at the lower levels. Thus, every level is in a sense, just as real as every other, since the ``whole picture'' cannot be deduced by starting at the “lowest level” and working upward'' (\cite{Talbot:2017}, p.\ 246). 

Similar concepts are expressed in a letter sent in early 1952 to Hanna Loewy. Here we read even more clearly about the ontological dependence relations existing between levels as well as the reasons for which a reduction to a fundamental ontology would be untenable:
\begin{quote}
all matter contains an infinity of qualitatively different levels, all interconnected. Moreover, there is another interesting point. The so called ``particles'' of any given level are made up of structures in the ``particles'' of the lower level, etc. ad infinitum. [...] Because of the infinity of levels, you cannot say that there are any ultimate ``individuals'', which are ``fundamental'' in the sense that their character is unalterable, and their existence eternal. At any level, any particular form of matter can always come into existence \& go out of existence as a result of a transformation in the components existing at a lower level, but only matter as a whole, in its infinity of properties and possibilities, is eternal. [...]  I should also add that I believe that no law is absolute or final, but that each law provides a successively better approximation to an absolute truth, that we can never possess in a finite time, because it is infinite in all its aspects, both qualitative and quantitative (\cite{Talbot:2017}, pp.\ 123-124).
\end{quote}

\noindent This letter is important for the purposes of the present paper, since it provides further evidence that Bohm endorsed a form of scientific realism, for according to him as science progresses better approximations to an absolute or noumenal truth about reality are discovered, although a full comprehension of it will never be achieved. Moreover, from this quote one can understand that in spite of physical theories being adequate only within a certain domain of application, they should not be considered false, but rather relatively or partially true---as we have previously seen discussing the limited applicability of Newtonian mechanics. Notably, given that for Bohm every theory is only applicable in certain specific range of energy/length scales, each framework can be only partially true. Consequently, one should not claim that classical mechanics is wrong because it has been proved inadequate to describe empirical phenomena at microscopic regimes, but rather its inability to represent certain experimental facts tells us the limits in which such a theory is (or is not) a correct description of the world.

Referring to this, another letter sent to Yevick on 31 March 1952 is relevant to clarify a further aspect of infinitism. In this correspondence one deduces that she tried to understand the latter in terms of Cantor's theory of transfinite numbers. Bohm's reply underlines that while Cantor's infinities consist in a collection of discrete, separated individuals all similar to each other, the levels of reality he is speaking about are all qualitatively distinct and different, so that each one must be treated independently with respect to the others.
More importantly, says Bohm, his view of reality would avoid the mechanistic and deterministic philosophy of XIX century physics as well as the a-causal metaphysics of quantum theory. On the one hand, the infinite number of levels entails by construction that nature cannot be explained and reduced to a finite number of fundamental entities and laws---against the mechanistic materialist spirit of the pre-quantum era. On the other hand, a causal description of physical phenomena can be retained also at the quantum (and sub-quantum) domain as explicitly show in his 1952 papers. In a nutshell, as Bohm wrote, ``although each level is causal, the totality of levels cannot ever be taken into account. Thus, as a matter of principle, we say that complete determinism could not even be conceived of, yet, each level can be determined'' (\cite{Talbot:2017}, p.\ 254). To this specific regard, in one of the final passages of the letter he explained the difference between causality and determinism, two notions that although tightly related are not equivalent. The former, to be understood as efficient causality at this stage of Bohm's career, entails that knowing the cause of a certain fact, we know that its effects will be obtained. Conversely,  if we manipulate and modify the causes, one thereby changes the effects ``in a predictable way''. On the other hand, the latter notion implies only predictability, but not the possibility of changing initial conditions. Assuming that reality can be described in a finite number of levels, then causality would be equivalent to determinism---i.e.\ the future would be logically contained in the present, writes Bohm. On the contrary, by stipulating the existence of an infinite number of layers we cannot in principle ``conceive the world as completely determined''. 

These ideas continued to be developed in the following years, where it became even more evident that infinitism was essential to avoid a completely deterministic and mechanistic perspective, according to which the world would be reducible to a set of basic primary entities. In the following letter we have still other demonstration that Bohm's project was not to restore a deterministic world-view against what Pauli, Rosenfeld and others erroneously thought. Interestingly, in his correspondence with Melba Phillips dated 15 March 1954 he explained the logical relations between causality and a mechanistic worldview:
\begin{quote}
it is necessary to sharpen the distinction between causality and mechanism (or deterministic mechanism). Mechanism is characterized by two fundamental aspects: (1) Everything is made of certain basic elements which themselves never change in essence (i.e., qualitatively). (2) All that these elements can do is to undergo some quantitative change according to some fixed laws of change. For example, if they are bodies, they can move in space. If they are fields, they can change their numerical values, etc. But the basic elements themselves never undergo qualitative change. If we postulate an infinity of levels, then we make a step beyond mechanism. For the elements existing at each level are made of still smaller elements in motion (i.e., changing quantitatively), and the mode of being of the higher level elements arises out of the motions of the lower level elements. Thus, there are no elements that can never change. Indeed, even if we have a finite number of levels, some qualitative change is possible within a mechanistic theory. For example, with atoms in chaotic motion, we obtain new large scale properties, such as pressure, temperature, etc., new entities, such as gas, liquid, solid, and qualitative changes between them. Now, at first sight, it may seem that we could eliminate the large-scale level by analyzing it in terms of its basic molecular motions. And if there were a finite number of levels, this would be true. But if there are an infinite number, then each level stands on a footing that is, in the long run, as basic as that of any other. For every level has below it a deeper one. Indeed, matter can be regarded as made up of the totality of all levels'' (\cite{Talbot:2017}, p.\ 170).
\end{quote}

From this quote we can infer that any layer of reality should be treated independently of any other, although dependence relations do exist, as we have already underlined in this section. Notably, the laws of a specific given theory are insensitive to the motions of more fundamental levels. This emphasizes the pluralistic ontological views supported by Bohm (cf.\ the next section). To this specific regard, Bohm claims that albeit often one may infer many features of a certain set of objects studying the behaviour of its components, there are cases in which ``there may be properties that cannot so be deduced. Not only may these properties be peculiar to a given level, but they may involve “crossing” of levels. For example, the general large scale conditions, such as electric field, gravitational field, may actually change the conditions of existence of smaller particles, such as electrons, neutrons, etc., so that in strong fields, the very “elementary” particles into which we now analyze matter would change. Thus, there can be a reciprocal influence from a higher to a lower level, which by itself would make impossible a complete analysis of all properties of the higher level in terms of the lower'' (\emph{ibid.}). Hence, Bohm concludes, each level contributes in its own way to the totality of reality.

To close this section it is interesting to note that Bohm discussed his views about the infinite richness of nature also with Einstein. Let us quickly contextualize their exchange, since it provides a nice summary of Bohm's views about quantum theory and the structure of reality at the end of 1954. 

In a letter dated 28 October 1954 Einstein wrote to Bohm that every effort so far made to complete quantum theory was not satisfactory, even his own attempt at generalizing the law of gravitation including the atomistic and discrete nature quantum systems.\footnote{Einstein discussed at length with Bohm the merits and problems of the causal interpretation. However, he did not find the pilot-wave theory an adequate solution to the problems of QM. For details cf.\ \cite{Einstein:1953} and \cite{Myrvold:2003}.} He closes by saying that if a fundamental ontology of fields cannot be given as foundations for an objective description of reality, then the notion of continuum should be abandoned altogether even in the context of space and time. 

On 14 November 1954 Bohm replied to Einstein with a long letter. In the first place, he holds a different position with respect to the father of relativity, claiming that the possibilities of an objective description of nature in terms of continuous notions have not been all analyzed and exhausted. In particular, he claims that studying the macroscopic structure of reality---in this case starting from general relativistic considerations---brings little advantage in order to find the correct laws for the microscopic regimes. In fact, while the microscopic structure of the world is only very weakly reflected at the macroscopic regimes, analyzing the microscopic structure one obtains greater insights about large scale phenomena and laws, which often are statistical approximations of the microscopic ones. Thus, he was doubtful that looking for the correct quantum laws starting from macroscopic field laws would have been a fruitful methodology. The first part of the letter is interesting for us since it provides useful insights to better characterize Bohm's thoughts about the structure of reality. Indeed, although our universe is composed by an infinite number of levels, there are relations among them. More precisely, less fundamental levels are ontologically dependent on more fundamental ones, although they are not strictly speaking reducible to them, because of the infinite chain of layers. In addition, Bohm explicitly affirms that macroscopic laws, i.e.\ roughly speaking those of classical physics dealing with larger scales, are statistical approximations of microscopic structures and laws.

In the second place, but related to the first point, Bohm affirms once again that in his views there would be another sub-quantum level beyond QM characterized by a continuous and causally determined motion. The basic entities of such a sub-quantum level would be a set of fields obeying non-linear equations; this field needs not be defined as other classical fields, rather the latter would emerge as averages from the motion of this deeper entity as well as the usual quantum mechanical wave function. What happens at the quantum level, he conjectures, would be determined by the evolution of a yet-unknown but qualitatively novel set of entities. In particular, the Schr\"odinger equation would describe averages of the dynamical effects of sub-quantum entities. More specifically, the relation between the sub-quantum and quantum scales can be associated to the relation between Brownian motion and ``the atomic level. In other words, events at the atomic level are contingent on the general irregular motions of some as yet unknown but qualitatively new kind of entity, existing below the atomic level''. In this context the $\psi$ function is to be conceived as an average of the dynamical motion of the lower level fields, and  assuming that ``the basic fields undergo a rapid, quasi-ergodic type of fluctuation, then with reasonable assumptions about these fluctuations, one obtains the Schr\"odinger equation as an average equation, satisfied by a suitable mean of the $\psi$ function''.

Interestingly, Bohm claims that although quantum theory would emerge as a statistical average from this deeper level, its laws will be insensitive with respect to the precise forms of the more fundamental dynamical equations valid at the sub-quantum level, highlighting thereby the autonomy of the quantum level with respect the more fundamental sub-quantum one---and consequently, generalizing this argument, the autonomy of each level. This fact in turn would show that it is very unlikely that we will be able to deduce of infer the motion and the behavior of the more fundamental fields from the non-relativistic quantum regime, since what is relevant at quantum scales are averages of sub-quantum dynamical evolutions. On the contrary, Bohm underlines that as soon as one has knowledge of the lower level, then one can derive more easily conclusions about the quantum level. Concluding the letter, Bohm eventually expresses his infinitistic view about reality\footnote{From Einstein's reply dated November 24, 1954 we understand that he did not shared Bohm's views about ``an unending hierarchy'' of structures and laws preferring a methodology based on logically simple laws with general validity.}:

\begin{quote}
On the whole, I do not find the idea of avoiding the continuum of space and time very plausible. I think rather that the continuum is infinitely rich in qualities. In other words, below any given type of level will always be a new level of motion and structure, so that each type of entity contains within it new types of entities that are still smaller. In general one may expect that irregular quasi-ergodic motion is characteristic of all the levels. Thus, every level will be subject to chance for fluctuations arising from the lower level motions. Nevertheless, there will be no limit to the application of causality, and to the possibility of making an objective description of these various levels of motions and of being. (Courtesy of the Birkbeck College Archives).
\end{quote}

As we can learn from the material studied in this section, Bohm continued to develop his ideas on infinitism during the years 1951-1954. We can then consider the reply to Takabayasi's paper as well as the letters to Yevick, Loewy, Phillips and Einstein as a portion of the preparatory work for the book \emph{Causality and Chance in Modern Physics}, which contains the most detailed illustration of Bohm's infinitistic views about the inherent structure of reality.
\begin{figure}
\includegraphics[scale=.65]{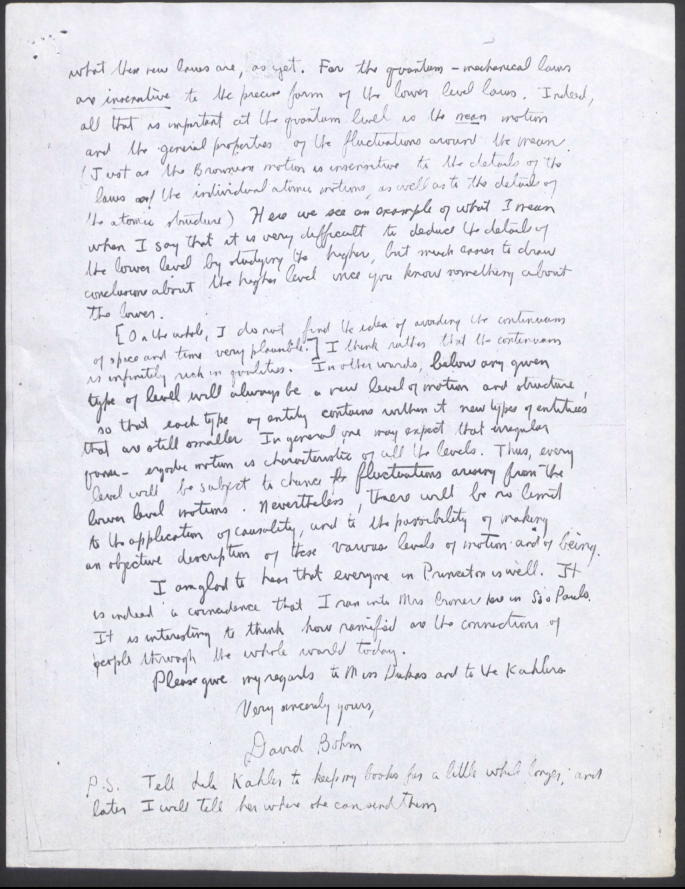}
\caption{This is the third page of Bohm's letter to Einstein dated 14 November 1954 containing a paragraph on the infinite structure of reality. Courtesy of the Birkbeck College Archives.}
\end{figure}
\clearpage 

\subsection{Causality and Chance in Modern Physics}

In 1957 Bohm published \emph{Causality and Chance in Modern Physics} which can be considered one of the pillars of his scientific and philosophical production.\footnote{From now on we will refer to this book simply as \emph{Causality and Chance}.} For it perfectly synthesizes his metaphysical reflections about the structure of reality, scientific theories, laws and causality, as well as it provides the basis to understand future directions that Bohm will explore in later years, as for instance the the philosophy of processes, developed mainly in 1960s with the volume \cite{Bohm:1965}, and holism, initiated in the early 1970s and culminated in the books \cite{Bohm:1980} and \cite{Bohm:1993aa}.  

\emph{Causality and Chance} begins with a discussion of causality which serves as a foundation for every other thesis defended in the book. Interestingly, he states a general principle---which is even prior to the notion of causality---which says that ``everything comes from other things and gives rise to other things'' (\cite{Bohm:1957}, p.\ 1). This statement is inferred from the empirical observations that in nature nothing remains constant, but rather everything is in perpetual modification and transformation, coming from something that existed before. Such a principle then prepares the basis for the notion of causality. In the study of nature, in fact, scientists found (and continue to find) patterns of connections and relationships between events, objects and phenomena which are \emph{necessary} in the following sense: (i) every time we observe the fact $A$, the effect $B$ will follow, and (ii) if we modify $A$, the effect $B$ will be modified accordingly in a predictable way. This is in essence the notion of causality at play in the book. 

Discussing a wide examples of causal relationships varying from medicine to theoretical physics, Bohm claims that since we can make predictions from them, these are neither the results of random connections between events, nor we can change them at our will, so that they represent objective features of our world. Moreover, in this philosophical framework such necessary or constant relationships define the way things are and behave, and are labeled \emph{causal laws}. He writes for instance: ``[t]he fact that such predictions are possible shows that the causal laws are not like externally imposed legal restrictions that, so to speak, merely limit the course of events to certain prescribed paths, but that, rather, they are inherent and essential aspects of these things. Thus, the qualitative causal relationship that water becomes ice when cooled and steam when heated is a basic part of the essential properties of the liquid, without which it could not be water. Similarly, the chemical law that hydrogen and oxygen combine to form water is a basic property of the gases hydrogen and oxygen, without which they could not be hydrogen and oxygen''. Analogously, he claims that since causal laws are essential in order to define the qualities and feature of physical objects, it would not be possible for us to conceive or discover them if they would not satisfy some sort of nomological regularity and objectivity. Alternatively, according to Bohm the simple fact that a certain object has a given attribute implies that it  ``will react in a certain way when it is subjected to specified conditions (e.g. the red object exposed to white light will reflect mostly red light). In other words, the causal laws that a thing satisfies constitute a fundamental and inseparable aspect of its mode of being'' (\cite{Bohm:1957}, p.\ 10). 

Interestingly, Bohm crucially underlines that in defining a certain law one must restrict the description of such a causal connection to the relevant factors which are necessarily involved in the problem under consideration, leaving outside the infinitely many other elements having negligible effects on it, because one may find an infinite number of contributing causes. Thus, the significant or relevant ones for a given effect are those that have appreciable influence on it in a certain context of interest. In particular, Bohm stresses that one always deals with incomplete precision while facing a scientific problem because every single event, process or phenomenon \emph{de facto} depends on an infinite series of elements. However, the majority of them do not pertain to the context under consideration and can be neglected. These negligible contributions will be then cancelled out and will not produce appreciable effects, so that it is actually possible to study a problem with a rigorous approximation, and without taking into consideration the actual infinity of factors which would be needed in order to obtain a completely perfect description of a phenomenon or for the prediction of a certain result. In turn, this entails that such perfect representations of physical phenomena are not achievable, and that our laws are always approximated descriptions of the physical reality, i.e.\ abstractions of real processes taking place in the world.\footnote{For more details of the various kinds of causal relations the reader may refer to \cite{Bohm:1957}, Chapter 1, Sections 7 and 8. Recent discussions that clarify Bohm's notion of causality and its relationship with determinism are contained in \cite{DelSanto:2023} and \cite{vanStrien:2024}.}

From Bohm's analysis of causal laws, hence, we can deduce the following conclusions that are essential to understand his philosophical views:
\begin{itemize}
\item Causal laws individuate constant relationships that represent objective features of reality in given domains of applications;
\item Such laws are fundamental in order to define properties of physical objects at the relevant scales. This entails that a law cannot be employed to characterize the attributes of items appearing in more fundamental theories, where new types of nomological relations occur; 
\item Similarly, they provide approximately correct descriptions of a certain set of events, processes and phenomena within a specific range of energy/length scales. As Bohm underlines, consequently ``any theory extrapolated to an arbitrary context and to arbitrary conditions will (...) lead to erroneous predictions. \emph{The finding of such errors is one of the most important means of making progress in science}. A new theory, to which the discovery of such errors will eventually give rise, does not, however, invalidate the older theories. Rather, by permitting the treatment of a broader domain of phenomena, it corrects the older theories in the domain
in which they are inadequate and, in so doing, it helps define the conditions under which they are valid'' (\cite{Bohm:1957}, p.\ 21, emphasis in the original);
\item Consequently, causal laws do not represent absolute truths because they cannot be applied universally, i.e. without approximations to every specific context at every scale. Thus, the goal of science is to find laws and theories which are progressively more fundamental and accurate.
\end{itemize}

Interestingly, Bohm presents his own metaphysical approach in contrast to the mechanistic philosophy usually attached to classical physics. Indeed, the second chapter of the book is devoted to the exposition of the main tenets of such a metaphysical account. These can be simply summarized saying that every portion of reality can be reduced to a finite set of absolutely fundamental laws and objects, so that every other physical entity, event and phenomenon can be explained in terms of these basic constituents. Hence, no new qualitative features of matter can arise from nature's elementary ingredients, that determine and exhaust all the possible modes of being. In particular, the philosophy of mechanism has been extrapolated from Newton's theory of mechanics (mainly by Laplace) assuming its universal applicability to every domain and layer of reality. Contrary to this view, Bohm argues that Newton's laws of motion form the formal basis of the science of mechanics, which by itself \emph{does not provide} a complete determination of the future behaviour of the whole universe. The premise that everything must be described and fall into the domain of application of Newtonian mechanics, then, represents a mere projection of Newton's theory to all possible contexts and domains of applications. However, this generalization is not grounded in scientific facts, rather it is the consequence of a philosophical edifice that conceived our universe as a mechanism built upon a restricted set of entities and laws. Therefore, Bohm claims that ``mechanism cannot be a characteristic of any theory, but rather, as we have already stated above, a philosophical attitude towards that theory. Thus, it would have no meaning to say, for example, that Newtonian mechanics is mechanistic; but it has meaning to say that a particular scientist (e.g.\ Laplace) has adopted a mechanistic attitude towards this theory'' (\cite{Bohm:1957}, p.\ 25).

In the fifth and last chapter of \emph{Causality and Chance}, we find a detailed tripartite objection against mechanistic philosophy. In the first place, history of physics disconfirms the basic tenets of this view, since the revolutions that occurred from Newton to this day radically changed the structure of physical theories, introducing entities and laws in open contrast with respect to those of classical physics. Moreover, Bohm notably argued that in virtue of the crisis that physics was facing after the Second World War---in particular with several issues affecting quantum field theory---future theoretical frameworks would have been as revolutionary as QM was compared to classical mechanics. In the second place, the assumptions concerning the final character of any particular ontology are neither necessary, nor empirically provable, because future theories may demonstrate their limited validity. Referring to this, Bohm exemplifies his argument taking into account the transition between classical and quantum mechanics, pointing out that ``Newton's laws of motion, regarded as absolute and final for over two hundred years, were eventually found to have a limited domain of validity, these limits having finally been expressed with the aid of the quantum theory and the theory of relativity'' (\cite{Bohm:1957}, p.\ 90). Finally, the mechanistic philosophy contravenes the principles of the scientific method, since the latter imposes that every object and law must be continuously subjected to verification. This process of testing may well end up in finding a contradiction with new discoveries or new domains of science. Looking at how physics evolved, claims Bohm, such contradictions not only systematically appeared, but also led to a deeper comprehension of the world. 
 
Contrary to the mechanistic philosophy, he proposed a version of metaphysical infinitism according to which there is no a bottom ground of fundamental entities and laws upon which everything else depends. As we have seen in the previous section, an important feature of Bohm's infinitism consists in rejecting the universality of any known ontology and set of laws, i.e.\ to reject the idea that a certain class of objects and causal relations can be successfully and perfectly applied to every level of reality. In essence, Bohm stated that looking at (i) the historical evolution of physical sciences, (ii) the available experimental data as well as (iii) the then-current crisis of theoretical physics that was shaking the foundations of the quantum theory of fields, one is pushed to endorse a conception of nature constituted by an infinity of different entities and causal relations (\cite{Bohm:1957}, p.\ 91). According to this view of science, physical theories do not always lead us closer to a fundamental ground, but instead show the infinite complexity of our universe. Furthermore, he believes that empirical data cannot a priori provide any justification to metaphysical restrictions concerning a particular set of items to be chosen as absolutely ontologically independent.\ On the contrary, conforming to Bohm's infinitism, scientific practice always discloses new entities, laws and phenomena which contribute to our continuous, never-ending process of understanding the limitless structure of reality. 

However, although Bohm denied the existence of a fundamental level, he firmly believed that every theory must be ontologically unambiguous in its domain of application. Therefore, theoretical frameworks must provide a clear ontology to be applied at the relevant energy/length scale, meaning that (i) the terms appearing in the vocabulary of a physical theories should  refer to/denote objects existing in the world, and (ii) the entities forming the basic ontology of a given theory have to be considered \emph{relatively} fundamental. Referring to this, he stated that 
\begin{quote}
[a]ny given set of qualities and properties of matter and categories of laws that are expressed in terms of these qualities and properties is in general applicable only within limited contexts, over limited ranges of conditions and to limited degrees of approximation, these limits being subject to better and better determination with the aid of further scientific research (\cite{Bohm:1957}, p.\ 91).
\end{quote}

\noindent Thus, a well-defined physical theory should provide a clear ontological picture for the domain in which it is a reliable description of physical phenomena. Nonetheless, its ontology and laws can be substantially modified with the progress of scientific research. This certainly exemplifies Bohm's scientific pluralism and his heterodox metaphysical views with respect not only to the dominant paradigm towards the interpretation of QM (cf.\ \cite{vanStrien:2019}), but also to a reductionist conception of science. 

Related to this, Bohm acknowledges that infinitism is a metaphysical thesis that cannot receive direct empirical confirmation exactly as mechanicism. However, the former is more adherent to the scientific practice since it takes seriously into account the crucial role played by boundary conditions and approximations in setting the limits to the validity of each physical theory. Moreover, assuming an infinite layers of reality one may apply a mechanistic viewpoint in each level, avoiding nonetheless a strong reductionist perspective as well as absolute determinism, as we have already seen in several letters in the previous section. 

Speaking about reductionism, it is worth noting that although there are relations of ontological dependence among levels, it is possible to study and describe each layer independently to the others. Already in the second chapter of the book Bohm explained the autonomy of levels through examples taken from classical statistical mechanics. Indeed, it was pointed out that the kinetic theory of gases was one of the first examples in which large-scale, macroscopic regularities, albeit ontologically dependent upon the microscopic molecular structure of the gas, where independent respect to the precise details of the molecular motions occurring at the microscopic regime---in fact, a given macro-state can be multiply realized by an infinity of micro-states. To this precise regard, Bohm claimed that
\begin{quote}
macroscopic average quantities (such as the mean number of molecules in a given region of space or the mean pressure
on a given surface) are extremely insensitive to the precise motions and arrangements in space of the individual molecules. This insensitivity originates, at least in part, in the fact that an enormous number of different motions and arrangements in space can lead to practically the same values for these quantities. [...] Because these mean values depend almost entirely only on the general over-all properties of the molecules, such as the mean density, the mean kinetic energy, etc., \emph{which can be defined directly at the large-scale level, it becomes possible to obtain regular and predictable relationships involving the large-scale level alone}.
It is clear that one is justified in speaking of a \emph{macroscopic level} possessing a set of relatively autonomous qualities and satisfying a set of \emph{relatively autonomous relations} which effectively constitute a set of macroscopic causal laws (\cite{Bohm:1957}, p.\ 34).
\end{quote}

\noindent Examples of such autonomy of levels can be found very easily in physics, as for instance one may consider that non-relativistic quantum theory is insensitive to the internal structure of nuclei, which are instead studied at the level of the quantum theory of fields, or that the latter theory is insensitive to the events and phenomena taking place at Planck's scale, \emph{etc.}. Thus, one may fairly say that at the various levels of description one finds a relative autonomy of behavior, so that one can study a given set of entities, laws and relationships ``which are characteristic of the level in question'' (\emph{ibid.}). 

Referring to the relative independence of each layer of reality, Bohm interestingly claims that although every entity, process or phenomenon are dependent upon an actual infinity of other qualities and relations which are all interconnected---here we can see the seeds of Bohm holism which will become primary in later stages of his work---one of the essential problem of science is to practically disentangle such causal relationships in order to be capable of dealing with subsystems of the universe and particular set of causal laws. In fact, he continues, it is a crucial task for scientists to individuate those entities and laws in a given level which are able to influence other things without themselves being significantly influenced. For instance, taking into account the example of the kinetic theory of gases considered a few lines above, one may say that despite the tight connections between macroscopic and microscopic states, the former possess a relative autonomy in their modes of being, therefore, one can study their features and behavior independently of their internal microscopic structure.

One of the most philosophically significant aspects of Bohm's views emerges from two contrasting features of his view that we just mentioned, namely the relative autonomy of levels of description on the one hand, and the actual dependence of physical objects and processes upon an infinity of other entities and causal relations on the other hand. From this apparent tension he remarkably concluded that the notion of ``thing'' or ``object'' is an idealization and an abstraction from the infinite background of structures that provide a given entity its conditions of existence. Therefore, he affirms that 
\begin{quote}
the notion of the infinity of nature leads us to regard each thing that is found in nature as some kind of abstraction and approximation. It is clear that we must utilize such abstractions and approximations if only because we cannot hope to deal directly with the qualitative and quantitative infinity of the universe. The task of science is, then, to find the right kind of things that should be abstracted from the world for the correct treatment of problems in various contexts and sets of conditions. The proof that any particular kinds of things are the right ones for a given context is then obtained by showing that they provide us with a good approximation to the essential features of reality in the context of interest (\cite{Bohm:1957}, p.\ 100).
\end{quote}

Given the practical and conceptual impossibility to deal with the infinite complexity of reality, Bohm believes that physical theories should be considered abstractions from the actual structure of our universe, capturing only the relevant and essential qualities and processes with which we can achieve a faithful---but always approximated---description and knowledge of nature. With the progress of science, then, we will achieve increasingly better representations of matter, although a one-to-one correspondence between our physical theories and reality will never been obtained.

This conclusion bring us to analyze the last section of \emph{Causality and Chance} dealing with the notions of truth and objective reality, which will be important for the reminder of this essay. There Bohm drawn the logical conclusions from all his previous arguments. Given that reality is composed by an infinite multiplicity of layers, whose entities, laws and relations are all reciprocally interconnected, and given that such levels are relatively autonomous, each theoretical framework will be able to uncover only relative truths, valid at certain levels but never universally applicable and/or generalizable. Nonetheless, as already underlined in this section, physical laws do represent objective aspects of reality, for they describe necessary connections between entities, events, phenomena, etc., which are independent of our minds, wills or ``the way in which we think about things''. 

\section{The Internal Realist View}
\label{IR}

As we have seen several times in the previous sections, Bohm often provided evidence for his arguments from the history of physics, as it would have become customary in philosophy of science in the following years. The mechanistic view is in fact criticized for its unwarranted universalization of Newtonian mechanics, extending its application to every scale. Contrary to this philosophical perspective, Bohm emphasized that so far every physical theory has been shown to be a reliable description of only \emph{some limited portion} of our universe, i.e.\ valid only within a specific interval of energy/length scales. Hence, theoretical entities and laws describing a certain layer of it are only relatively fundamental---i.e.\ given that each level can be studied independently, its objects and dynamical laws will be considered fundamental only \emph{at that particular level}. Therefore, one inductively infers that future theories will be only valid within certain specific domains of application. Moreover, because history of physics shows that scientific investigations always disclosed new features and levels of reality, he claimed that its structure is limitless and not reducible to an absolute bottom level. Hence, we will be unable to embrace the infinite complexity of nature with a finite set of entities and laws, so that no final theory of everything will be possibly found. Let us call Bohm's argument the \emph{infinitistic meta-induction}.

According to this meta-induction, scientific theories progressively come closer to a true description of reality with increasingly better approximations, expanding our knowledge of the world by discovering new layers of nature. It is worth recalling once more that Bohm was a metaphysical realist, namely he thought that the external world exists independently on our minds, knowledge and possible observations of it. Indeed, reality is explicitly defined in \emph{Causality and Chance} as the totality of existing matter, laws and relations in their continue process of becoming (cf.\ \cite{Bohm:1957}, pp.\ 114-115). However, given the ideas expressed in the latter book is not trivial to claim that Bohm was a full-fledged scientific realist, since he did \emph{not} believe that the sentences or statements expressed by scientific theories are \emph{literally true}, as we have already pointed out in the previous section. In particular, one must take into account that our best theoretical frameworks are abstractions, i.e.\ idealized representations of the real objects and causal relationships actually constituting our universe. Let us then try to understand what kind of scientific realism best approximates David Bohm's perspective, relying on the main philosophical conclusions individuated in the writings analyzed so far.

In the opinion of the present author one may interpret his approach as a form of internal realism, according to which it is possible to maintain a realist ontological commitment towards the theoretical entities and laws appearing in our current physical theories, without being committed to a fundamental, scale-invariant ontology. According to such an account, given a particular theory $T$, one would be uniquely committed to the existence of those entities and laws in $T$ with a direct physical meaning, however, such commitment would be constrained by $T$'s specific domain of application.

For instance, if the non-relativistic quantum regime would be correctly described by the pilot-wave theory, one would be ontologically committed to the presence in the world of quantum particles with certain definite properties, e.g.\ position and velocity, as well as of a new kind of field, namely the $\psi$-function. Considering instead classical electromagnetism or general relativity as valid representations of other levels of reality, one would accept the actuality of different kinds of physical fields. Nonetheless, existential claims implying the reality of entities contained in the vocabularies of these theories can be considered approximately true only within their respective domains of validity. Consequently, the ontological commitment implied by the mentioned frameworks is limited to the specific length scales in which they can be considered approximately reliable descriptions of nature. 

Moreover, Bohm argued that the ontology of a certain theory may be subjected to substantial modifications with the progress of scientific research. For example he underlined several times that the metaphysical content of the causal interpretation may vary importantly at the sub-quantum level. Similarly, Maxwell and Einstein theories cannot be extended to more fundamental domains, e.g.\ to the quantum regimes, without substantial modifications of their metaphysical content and laws. Hence, according to this form of internal realism, the fundamentality of a given ontology will be always relative to the particular theory at hand and bounded by its limited range of application. This feature portraits accurately Bohm's belief, since he stated that any given set of entities and laws are in general applicable only within ``limited contexts, over limited ranges of conditions and to limited degrees of approximation'', as we have already seen. 

It is worth noting that endorsing this internal realist view in the context of metaphysical infinitism, one has to accept the possibility to have ontological discontinuity between physical theories: the entities and laws defining a certain framework may in fact greatly differ with respect to those employed at deeper or less deep scales. This fact however caused no troubles for Bohm, who explicitly admitted the presence of contradictory types of motion at different levels:  
\begin{quote}
the relative autonomy in the modes of being of different things implies a certain independence of these things, and this in turn implies that contradictions between these things can arise. For if things were co-ordinated in such a way that they could not come into contradiction with each other, they could not be really independent. We conclude, then, that opposing and contradictory motions are the rule throughout the universe, and this is an essential aspect of the very mode of things (\cite{Bohm:1957}, p.\ 102).
\end{quote}
\noindent Such views certainly exemplify Bohm's pluralist views concerning the ontology of physical theories; after all, according to his perspective one cannot resort to a monist metaphysics in order to describe the infinite richness and complexity of nature (cf.\ also \cite{vanStrien:2019}). Hence, following internal realism as formulated by Bohm one may conclude that (i) whatever ontology works at the certain regime, it can be modified at more/less fundamental levels, and (ii) ontological inconsistency between theories defined at diverse energy/length scales can be tolerated, given the provisional and fallible character of scientific knowledge.

Referring to this, it should be underlined that the possibility to have different ontological commitments at diverse energy/length scales is not a negative consequence of Bohm's proposal, but rather an advantage to understand the complexity of reality as well as of contemporary science. Indeed, many philosophers in the recent years proposed pluralist views in order to grasp the different notions at play in various physical theories. Interesting examples are given by \cite{Lombardi:2005} and \cite{Scerri:2007} who argue for the ontological independence of chemistry with respect to quantum physics. Moreover, analyzing carefully the historical evolution of QM and the measurement techniques in particle physics, \cite{Falkenburg:2007aa}, p.\ 38 observes that
\begin{quote}
the unity of physics is a \emph{semantic} rather than an ontological unity. Physics still has a unified language, namely the language of physical quantities, even though the unity of axiomatic theories and their objects has been lost.
\end{quote}

\noindent Taking into account the notions of e.g.\ ``particle'' or ``field'', one cannot but note that they have complete different meanings when included in classical or quantum theories. Similar considerations can be made about the concept of spacetime. Referring to the tension existing between realism about spacetime in general relativity and a functionalist perspective in quantum gravity, Lam \& W\"utrich interestingly claim that a possible way to resolve it is to appeal to a local interpretation of theories, whose commitments may also be divergent:
\begin{quote}
this piecemeal approach leads ourselves to incline towards a geometric structural realism about spacetime in GR, and spacetime functionalism in much of QG. Obviously, there is no requirement that these `locally optimal' interpretations are consistent across contexts. Taking naturalism seriously mandates local interpretations of theories, i.e., their reading needs to start from the theories themselves, rather than from a presupposed and fixed interpretative scheme or set of demands. Given that scientific revolutions may bring with themselves a shift in methods, aims, and values that constitute a scientific paradigm, naturalism prohibits an inflexible a priori commitment to a particular interpretative template. Thus, we believe that scientific realism tempered by naturalism must accept the possibility that our interpretative stances in GR and in QG diverge (\cite{Lam:2020}, pp.\ 349-350).
\end{quote}

Hence, the ontological unity hoped by a reductionist or mechanistic philosophy seems to be at odds with the current development of physics, making Bohm's philosophical reflections still relevant and interesting for the present day discussion about scientific realism.\footnote{\cite{Lam:2020} make an interesting case arguing that naturalism leads to a ``local''---or in our wording ``internal''---interpretation of physical theories.}$^,$\footnote{Referring to this lack of unity, it should be noted that one may frame internal realism in a \emph{moderate} pluralist account, which contemplates the possibility that the ontological plurality currently present in contemporary physics will be resolved in the future. Indeed, conforming to moderate pluralism, the final goal of every scientific domain is to establish a unique and complete account of the natural phenomena lying within its scope, making it also compatible with other accounts of other scientific domains. Thus, according to this moderate view, different domains of physics may be integrated and synthesized without ontological inconsistencies in future developments of the discipline (cf.\ \cite{Seselja:2017}). This project would however imply the rejection of infinitism, hence, it would not be completely adherent to Bohm's view.} Moreover, it is interesting to note that in the literature concerning scientific pluralism, several philosophers of science claim that incompatible or inconsistent theories can be simultaneously accepted since it is methodologically fruitful for scientific inquiry, as Bohm repeatedly stressed---cf.\ for instance \cite{daCosta:2003}, \cite{Rescher:1988} or \cite{Ruphy:2011}. Indeed, da Costa and French argue that the notion of ``belief'' used by scientists in accepting a given theoretical framework is not the philosophical concept of a true proposition, but a more vague notion reflecting epistemic fallibility. According to them, to accept a theory $T$ is not to believe that $T$ is true, but rather to reason \emph{as if it were true}. Similarly, Rescher claims that a reason to accept inconsistency among theories is provided by the fact that scientific reasoning is plausible but always fallible, thus, ``while consistency should play the role of a regulative ideal, it may at times be sacrificed for the sake of other cognitive values'' (\cite{Seselja:2017}, p.\ 12).

Finally, let us emphasize that such an internal realist view, although casted within a pluralist and infinitistic background, can be properly considered a genuine form of scientific realism. Indeed, it demands metaphysical clarity and explanatory power from physical theories: every framework---as for instance the pilot-wave theory, Newtonian mechanics, Maxwell's electromagnetic theory, \emph{etc.}---provides a finite set of entities and laws which is responsible for the explanations and predictions of the theory at hand within its well-defined domain of application. Alternatively stated, the theoretical machinery of a certain theory does have a \emph{functional} role for the explanation of a specific set of phenomena, and the entities appearing in its equations acquire a precise set of attributes only when inserted within a set of laws constraining their behaviour, as already underlined in the previous section (for more details cf.\ \cite{Bohm:1957}, Chapters 1 and 2). In addition, the present form of internal realism manifests the tridimensional character that is generally ascribed to scientific realism (cf.\ \cite{Chakravartty:2017}, Section 1.2). Firstly, it implements metaphysical realism, i.e.\ the idea that there exists an external world independently of human observers. Secondly, from a semantic perspective Bohm claimed that physical theories provide an approximately true description of natural phenomena, in the sense that scientific laws describe objective and necessary features of the world. The only limitation imposed by his version of internal realism is that such descriptions are reliable only within a definite domain of application, i.e.\ they are not universally valid. Finally, the explanations of physical phenomena given by scientific theories do provide knowledge of the external world. Therefore, we can safely conclude that the metaphysical, semantical and epistemological aspects of scientific realism are conserved within Bohm's philosophical reflections about the aims and scope of physical theories.

\section{Closing Remarks}
\label{conc}

In this essay we have argued that in order to explain the negative reception of David Bohm's pilot-wave theory one should take into due account a variety of factors, many of which are not strictly speaking scientific. Indeed, several objections against the causal interpretation stand on purely philosophical grounds, as in the cases of Rosenfeld and to a lesser extent Pauli. Moreover, the sociological and historical analyses mentioned in the introduction are key to understand the hostile attitude of many physicists who were reticent even to admit or to conceive the possibility of alternative formulations of quantum theory. A similar reception, unfortunately, affected also Hugh Everett's relative-state formulation of QM, as clearly illustrated in \cite{Freire:2015aa} and \cite{Osnaghi:2009}. 

In particular, studying the correspondence between Bohm and Pauli as well as the difficult interactions that the former had with Rosenfeld, we have seen how these great scientists misunderstood the main goals and motivations that led Bohm to propose a new reading of the quantum formalism. In fact, we showed that to restore determinism or to anchor physics to outdated ideas were not aims in his agenda. On the contrary, Bohm's letters and published works explicitly show an open-minded thinker who wanted to avoid absolute determinism and the mechanistic philosophy attached to Newtonian physics in the first place. Thus, Bohm's contemporaries evidently had all the necessary material and information not to classify or to consider him a reactionary scientist. 

Secondly, we reconstructed in some detail the evolution of Bohm's ideas about (i) the structure of reality, and (ii) the ontologico-epistmological role of physical theories from the early fifties until the publication of his monograph \emph{Causality and Chance in Modern Physics}. Reading his private and public production between 1951 and 1957 one cannot but note the originality and depth of his philosophy of science, which can surely be interesting for contemporary discussions about scientific realism and pluralism, as we have argued in the previous section. It is immediately clear in fact how accusations of dogmatism (still present today) miss completely the point of Bohm's scientific and philosophical research. 

In conclusion, let us underline that further work can be done in providing a coherent account of Bohm's complete philosophical trajectory, i.e.\ putting the ideas here presented in relation with later stages of his career, and thereby with his successive scientific productions. Moreover, it would be interesting to study the connections and relations existing between his philosophy of science and other important perspectives as for instance those advanced by Kuhn and Popper, whose work was very well known by Bohm.\footnote{Cf.\ \cite{vanStrien:2019} for an interesting discussion of Bohm's influence on Feyerabend.} Finally, it will be certain relevant to understand how the modern supporters of pilot-wave theory relate themselves to his heritage. All of this will be material for future research.

\bibliographystyle{apalike}
\bibliography{PhDthesis}
\clearpage
\end{document}